\documentclass[reprint, aps, superscriptaddress,preprintnumbers,showpacs, hyperref]{revtex4-1}

\usepackage{amssymb}
\usepackage{color}
\usepackage{amsthm}
\usepackage{graphicx}
\usepackage{dcolumn}
\usepackage{bm}
\usepackage{subfigure}
\usepackage{amssymb}
\usepackage{verbatim}
\usepackage{amscd}
\usepackage{amssymb}
\usepackage{amsmath, amsfonts}
\usepackage{setspace}
\usepackage[]{graphicx}        
\usepackage{amsthm}
\usepackage{natbib}
\usepackage{enumerate}
\newcommand{\quotes}[1]{``#1''}

\def \>{\rangle}
\def \<{\langle}
\def \gen{\ensuremath{\mathcal{G}}}
\def \viol{\ensuremath{\mathfrak{e}}}

\newcommand{\ket}[1]{| #1  \>}
\newcommand{\bra}[1]{ \< #1  |}
\newcommand{\proj}[1]{\ket{#1} \bra{#1} }

\newcommand{\paren}[1]{\left ( #1 \right )}

\newcommand{\myequation}[1]{\begin{eqnarray} #1 \end{eqnarray}}
\newcommand{\myequationn}[1]{\begin{align*} #1 \end{align*}}

\def \forall{\text{ for all }}

\theoremstyle{plain}

\theoremstyle{plain}

\theoremstyle{plain}
 
\theoremstyle{plain}

\theoremstyle{remark}

\theoremstyle{conjecture}

\theoremstyle{observation}

\theoremstyle{definition}

\theoremstyle{corollary}

\theoremstyle{definition}

\theoremstyle{definition}

\theoremstyle{assumption}

\theoremstyle{definition}

\theoremstyle{problem}

\theoremstyle{fact}

\begin{document}

\preprint{MIT-CTP 4274}

\title{Non-perturbative gadget for topological quantum codes}
\author{Samuel A. Ocko}
\affiliation{Department of  Physics, Massachusetts Institute of Technology, Cambridge, Massachusetts 02139, USA}
\author{Beni Yoshida}
\affiliation{Center for Theoretical Physics, Massachusetts Institute of Technology, Cambridge, Massachusetts 02139, USA}

\date{\today} 

\begin{abstract}
Many-body entangled systems, in particular topologically ordered spin systems proposed as resources for quantum information processing tasks, often involve highly non-local interaction terms. While one may approximate such systems through two-body interactions perturbatively, these approaches have a number of drawbacks in practice. Here, we propose a scheme to simulate many-body spin Hamiltonians with two-body Hamiltonians \emph{non-perturbatively}. Unlike previous approaches, our Hamiltonians are not only exactly solvable with exact ground state degeneracy, but also support completely localized quasi-particle excitations, which are ideal for quantum information processing tasks. Our construction is limited to simulating the toric code and quantum double models, but generalizations to other non-local spin Hamiltonians may be possible.
\end{abstract}
\maketitle

Many-body entanglement arising in strongly correlated systems is a very promising resource for realizing various ideas in quantum information, such as quantum communication and quantum computation. In particular, topologically ordered spin systems can be employed for reliable storage of quantum information inside the degenerate ground space~\cite{Kitaev97} and for fault-tolerant quantum computation with non-abelian anyonic excitations~\cite{Kitaev03}. These topological approaches may resolve many problems in quantum information science; qubits are encoded in many-body entangled states and are thus naturally protected from decoherence. 

Unfortunately, topologically ordered spin systems capable of quantum information processing are very difficult to realize physically. Many proposed topologically ordered spin Hamiltonians, such as the toric code, quantum double model~\cite{Kitaev03}, and string-net model~\cite{Levin06}, involve highly non-local interaction terms; this is a stark contrast to Hamiltonians which occur in nature, which have only geometrically local two-body interactions. Moreover, the resource systems above are known \emph{not} to be supported by any two-body Hamiltonian ~\cite{Nielsen06}.  

Many efforts have been made to construct two-body Hamiltonians which ``approximate'' non-local resource Hamiltonians. The most commonly used approach is to approximate target Hamiltonians through so-called ``perturbative gadgets''~\cite{Kempe05, Bravyi08b, Jordan08, Koenig10, Kargarian10, Brell11}. The central idea of perturbative gadgets is to design a two-body Hamiltonian whose leading perturbative contribution gives rise to the desired many-body Hamiltonian; unfortunately, most obtained two-body Hamiltonians are not exactly solvable, and their properties are hard to determine except for a few exactly solvable examples~\cite{Kitaev06b, Yao07}. In addition, the perturbative Hamiltonian only approximates the target Hamiltonian, and may give a very weak effective Hamiltonian with a rather small energy gap. Furthermore, quasi-particle excitations (energy eigenstates) arising in perturbative Hamiltonians cannot be created through completely localized manipulations of spins; excitations are always delocalized and the ground state degeneracy might be split for finite system sizes, resulting in fatal errors in practice. While a non-perturbative approach based on the PEPS formalism was developed for simulating the cluster state for measurement-based quantum computation~\cite{Chen09}, such an approach may not be applicable to degenerate systems with topological order. 

Here, we propose a scheme to simulate topologically ordered Hamiltonians through two-body interactions \emph{non-perturbatively}. Our scheme builds on previously established ideas in perturbative gadget studies, such as the use of hopping particles proposed by K\"{o}nig~\cite{Koenig10}, and the encoding of single particles into multiple particles used by Brell \emph{et al}~\cite{Brell11}. Combining these remarkable insights, we are able to construct the first topologically ordered spin system which satisfies the following; 1) The Hamiltonian has at most two-body, geometrically local interactions. 2) The Hamiltonian has exactly solvable ground states and low-energy excitations, and is provably gapped for all system sizes. 3) The ground space of the Hamiltonian is exactly connected to that of the target Hamiltonian through local unitary transformations, and anyonic excitations are completely localized.

For clarity of presentation, we illustrate the gadget construction for the toric code. A generalization to the quantum double model is also possible (see appendix \ref{sec:double}).

\begin{figure}[htb!]
\centering
\includegraphics[height=0.90\linewidth]{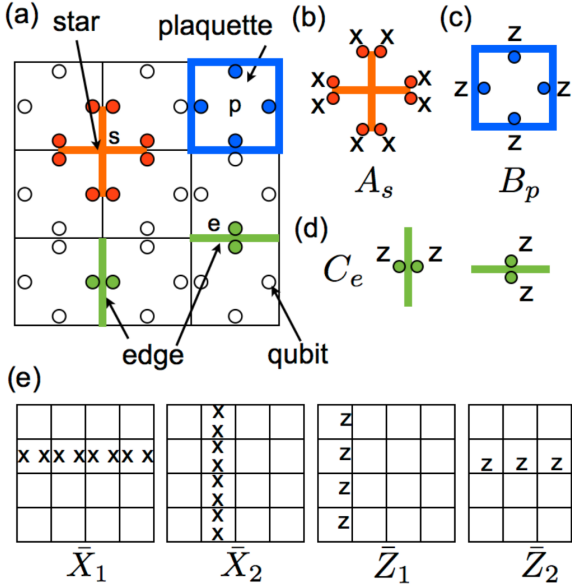}
\caption{(a) Construction of the modified toric code. Dots represent qubits. (b) A star term $A_{s}$ (red online). (c) A plaquette term $B_{p}$ (blue online). (d) An edge term $C_{e}$ (green online). (e) Two pairs of logical operators.
} 
\label{Toric}
\end{figure}

\emph{Modified toric code---}We begin by defining a modified version of the toric code, also known as the $Z_{2}$ lattice gauge model, that we will simulate through a two-body Hamiltonian. Consider a system of qubits defined on edges of a square lattice with periodic boundary conditions. Unlike the conventional toric code, \emph{two} qubits live on each edge in our construction (see Fig.~\ref{Toric}(a)), governed by the following Hamiltonian 
\begin{equation*}
\begin{split}
&H = - J \sum_{s} A_{s} - J\sum_{p} B_{p} - J\sum_{e} C_{e} \\
&A_{s} = \prod_{j \in s} X_{j}, \quad B_{p} = \prod_{j \in p} Z_{j}, \quad C_{e} = \prod_{j \in e} Z_{j},
\end{split}
\end{equation*}
where $s$, $p$ and $e$ represent ``star'', ``plaquette'' and ``edge'' respectively, as defined in Fig.~\ref{Toric}(b)(c)(d). $X_{j}$ and $Z_{j}$ are Pauli X and Z operators on qubit $j$, and $J$ is some positive constant. 
The model is exactly solvable since interaction terms commute with each other, and it can be considered to be a stabilizer code. The model has four degenerate ground states, as in the toric code. Inside of the ground space, $A_{s} = B_{p} = C_{e} = 1$, meaning $A_{s}|\psi \rangle = B_{p}|\psi\rangle = C_{e}|\psi\rangle = |\psi\rangle$ for all $s$, $p$ and $e$ when $\ket{\psi}$ is a ground state.  Notice that one can create the toric code from this model by applying controlled-NOT gates between pairs of qubits on each edge. Since the toric code and the modified model are connected through local unitary transformations, they are considered to be in the same quantum phase~\cite{Chen10, Beni10b}. The ground space of the modified toric code has a four-fold degeneracy, as seen by writing down two pairs of \quotes{logical operators} which commute with the Hamiltonian but anti-commute with each other (see Fig.~\ref{Toric}(e)). The non-locality of logical operators makes the model of great interest for robust storage of quantum information. 

As a first step towards obtaining a two-body Hamiltonian simulating this modified toric code, we group the four qubits in each plaquette into a single composite particle with a 16-dimensional space. While $B_{p}$ becomes one-body, and $C_{e}$ is two-body through this grouping, the star term $A_{s}$ is only reduced to four-body. Below, we provide a scheme to simulate $A_{s}$ through only two-body terms. 

\begin{figure}[htb!]
\centering
\includegraphics[width=0.75\linewidth]{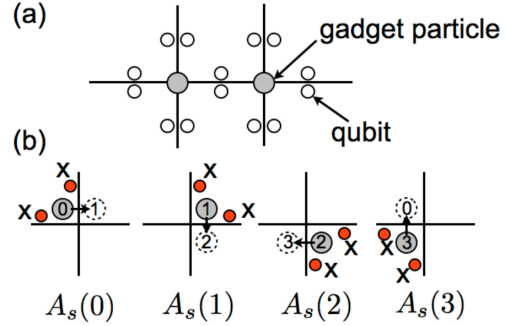}
\caption{Construction of the hopping term $D^{\dag}_{s}$. (a) Gadget particles at stars. (b) Terms $A_{s}(m)$ that are tensor products of two Pauli $X$ operators. Each term acts on two qubits (red online), depending on spin values of gadget particles. 
} 
\label{spin}
\end{figure}

\emph{Gadget Hamiltonian--} The central idea behind our construction is to add a \quotes{gadget particle} at each star (see Fig.~\ref{spin}(a)). The gadget particle has four possible spin values $m_{s}=0,1,2,3$. We replace the four-body star term $A_{s}$ with two-body terms $H_{hop}$ and $H_{shield}$ which involve the gadget particles:
\begin{equation}
\begin{split}
&H_{gadget} = H_{p} + H_{e} + H_{hop} + H_{shield}  \\ 
&H_{p} = - J\sum_{p} B_{p},\quad H_{e} = - J\sum_{e} C_{e}. \label{eq:gadget}
\end{split}
\end{equation}
The hopping term is $H_{hop}=\sum_{s}h_{s}$ where
\begin{equation}
\begin{split}
h_{s} &= -  U| m_{s}=0 \rangle\langle m_{s}=0|  -t \left( D^{\dagger}_{s}  + D_{s} \right)\\ 
D_{s}^{\dagger} &= \sum_{m_{s}=0,1,2,3} | m_{s}+1 \rangle \langle m_{s} | \otimes A_{s}(m_{s}) \quad \mbox{(mod $4$)} , \notag
\end{split}
\end{equation}
where $U$ and $t$ are some positive constants, and $m_{s}$ represents the spin value of the gadget particle at $s$. Terms $A_{s}(m)$ are products of two Pauli $X$ operators as depicted in Fig.~\ref{spin}(b). Since $A_{s}(m)$ are one-body operators when qubits in a plaquette are viewed as a composite particle, hopping terms are two-body. This hopping term will effectively induce star terms $A_{s}$ since $A_{s}= (D_{s}^{\dag})^4$.
The shielding term $H_{shield}$ consists of two-body interactions between gadget particles:
\begin{align}
H_{shield} =  J \sum_{s} T_{\ell}(m_{s})T_{r}(m_{s+\hat{x}}) + T_{d}(m_{s})T_{u}(m_{s+\hat{y}})  \notag
\end{align}
where $s+\hat{x}$ and $s + \hat{y}$ are unit translations of a ``star'' $s$ in the horizontal and vertical directions, and 
\begin{align*}
  T_{\ell}(m) &= 1-2 \delta_{m,2}, \quad T_{r}(m) = 2 \delta_{m,0} -1\\
T_{d}(m) &=1-2 \delta_{m,3}, \quad   T_{u}(m) = 1-2 \delta_{m,3}
\end{align*}
with $\delta_{m,m'}=1$ for $m=m'$ and $0$ otherwise. As we will see below, this choice of the shielding term decouples effective interactions between neighboring gadget particles, and makes the model exactly solvable.

\emph{Decomposition into subspaces---}Now, we solve the gadget Hamiltonian in Eq.~(\ref{eq:gadget}). It is convenient to decompose the entire Hilbert space into  subspaces. Let us denote computational basis states whose gadget values are all $|0\rangle$s: $|\psi(\vec{d})\rangle =  |\vec{0}\rangle_{gadget}  \otimes |\vec{d} \rangle_{qubit}$ where $|\vec{d}\rangle_{qubit}$ represents spin values $\ket{d_{j}}$ for qubits, and $|\vec{0}\rangle_{qubit}$ means that all the qubits are $|0\rangle$s. We define the subspace $\mathcal{M}(\vec{d})$ such that it is spanned by all the states which can be reached from $|\psi(\vec{d})\rangle$ by applying $D_{s}^{\dagger}$:
\begin{align}
\mathcal{M}(\vec{d}) = \big \langle \prod_{s} (D_{s}^{\dagger})^{\lambda_{s}}|\psi(\vec{d})\rangle, \ \forall \lambda_{s} \big \rangle.\notag
\end{align}

We can verify that $\mathcal{M}(\vec{d})$ is an invariant subspace of $H_{gadget}$. Then, one can solve the gadget Hamiltonian inside each subspace $\mathcal{M}(\vec{d})$ independently.

\emph{Ground state subspace---}
We will first solve for the ground state inside  $\mathcal{M}(\vec{0})$, and then will show its lowest energy state to be a ground state. We note that inside $\mathcal{M}(\vec{0})$ $B_{p} = 1$, and thus plaquette terms need not be considered. Denoting the total number of stars as $N$, we may view $\mathcal{M}(\vec{0})$ as the Hilbert space of $N$ particles. 
\begin{equation}
|\vec{\lambda}\rangle = \bigotimes_{s} \ket{\lambda_{s}} =  \prod_{s} (D_{s}^{\dagger})^{\lambda_{s}}|\psi(\vec{0})\rangle. \label{eq:lambda}
\end{equation}
Noting that  $(D_{s}^{\dagger})^{4} = A_{s},  \quad (D_{s}^{\dagger})^{8}=  I$, these particles can be considered to have eight-dimensional Hilbert spaces, $\lambda_{s} = 0, \ldots ,7$ \cite{OurFootnote}. In this ``$\lambda$-representation'', the hopping term $H_{hop}$ can be written as a \emph{one-body} Hamiltonian: $H_{hop}=\sum_{s}h_{s}$ where
\myequationn{
h_{s} =& - U \big( | \lambda_{s}=0  \rangle\langle \lambda_{s}=0| + | \lambda_{s}=4 \rangle\langle \lambda_{s}=4| \big)\\ 
&- t \sum_{\lambda_{s}=0}^{7} \big(|\lambda_{s} + 1 \rangle \langle \lambda_{s}|  + h.c \big) \qquad \text{ (mod $8$)}. \label{eq:hopping}
}
However, edge terms $C_{e}$ are not one-body inside $\mathcal{M}(\vec{0})$. 

A key idea behind our gadget arises from the fact that these two-body interactions arising from $C_{e}$ can be exactly cancelled by adding the shielding term $H_{shield}$. Inside $\mathcal{M}(\vec{0})$, edge terms have the same action as the following two-body terms involving gadget particles: $C_{e} =  T_{\ell}(m_{s})T_{r}(m_{s+\hat{x}})$ for a horizontal edge $e$ connecting $s$ and $s+\hat{x}$, and $C_{e} = T_{d}(m_{s})T_{u}(m_{s+\hat{y}})$ for a vertical edge $e$ connecting $s$ and $s+\hat{y}$, as one can verify from direct calculations (see appendix \ref{sec:shield}). Then, the edge terms are exactly cancelled: $H_{e} + H_{shield}=0$ inside $\mathcal{M}(\vec{0})$. This means the gadget Hamiltonian is one-body in the \quotes{$\lambda$-representation}: $H_{gadget}= \mbox{const}+\sum_{s}h_{s}$. 

Because of this, all energy eigenstates inside $\mathcal{M}(\vec{0})$ can be written in the tensor product form $\ket{\vec{\alpha}} = \bigotimes_{s} \ket{\alpha_{s}}$ where $\ket{\alpha_{s}} = \sum_{\lambda_{s}} \alpha_{s}(\lambda_s) \ket{\lambda_s}$. The lowest energy state is $\ket{\psi_{GS}(\vec{0})} = \bigotimes_{s}\ket{\alpha_0}$, where $\alpha_{0}(\lambda) = \alpha_{0} (\lambda +4) \forall \lambda$. Therefore, returning from the $\lambda$-representation, we can write the ground state as
\begin{equation*}
\begin{split}
|\psi_{GS}(\vec{0})\rangle &= \prod_{s}\sum_{\lambda=0}^{7}\alpha_{0}(\lambda)(D_{s}^{\dagger})^{\lambda}|\psi(\vec{0})\rangle \\
             &= \prod_{s}(I + A_{s})\sum_{\lambda=0}^{3}\alpha_{0}(\lambda)(D_{s}^{\dagger})^{\lambda}|\psi(\vec{0})\rangle.
\end{split}
\end{equation*}
We see that there is a finite energy gap inside $\mathcal{M}(\vec{0})$, since $H_{gadget}$ acts as a one-body Hamiltonian.

\emph{Unitary Connection---}This lowest energy state $|\psi_{GS}(\vec{0})\rangle$ is connected to the ground state of the modified toric code through the following local unitary transformation:
\begin{align}
U = \prod_{s} U_s, \quad U_{s} \equiv \sum_{m_{s}=0}^{3}| m_{s} \rangle\langle m_{s}|\prod_{m < m_{s}}A_{s}(m). \label{eq:unitary}
\end{align}
In particular, we have $U|\psi_{GS}(\vec{0})\rangle = |\tilde{\alpha_{0}} \rangle_{gadget}^{\otimes N} \otimes |\psi_{Toric}(\vec{0})\rangle_{qubit}$ where $\ket{\tilde{\alpha_{0}}}= \sum_{m=0}^{3}\alpha_{0}(m)\ket{m}$, and $|\psi_{Toric}(\vec{0})\rangle_{qubit} = \prod_{s}(I + A_{s}) |\vec{0}\rangle$ is a ground state of the modified toric code. We may verify that the gadget Hamiltonian has three other ground states $\ket{\psi_{GS}(\vec{d_{i}})}$, $i = 1,2,3$, inside $\mathcal{M}(\vec{d_i})$, connected in the same way to the ground states $\ket{\psi_{Toric}(\vec{d_{i}})}$ of the modified toric code. 

It is then simple to find the logical operators for the gadget Hamiltonian; they are those of the modified toric code conjugated by $U$: $U^{\dag}\bar{X}_{1}U$, $U^{\dag}\bar{X}_{2}U$ , $U^{\dag}\bar{Z}_{1}U$ and $U^{\dag} \bar{Z}_{2}U$. The ground space is topologically ordered since it meets the criteria for the stability against local perturbations proposed in~\cite{Bravyi10b}. 

Anyonic excitations, which are also energy eigenstates, can be created by applying \quotes{segments} of logical operators combined with local operations on gadget particles in a similar way to the conventional toric code. As a result, excitations can be created only through \emph{completely localized} manipulations of spins in small regions. This is in striking contrast to perturbative Hamiltonians where anyonic excitations are \emph{delocalized}, and cannot be created through completely localized manipulations of spins. 

\emph{Energy gap---}Finally, we show that $|\psi_{GS}(\vec{d_{i}})\rangle$ are the ground states of the gadget Hamiltonian. To do so, we prove that the lowest energy states within other non-ground-state subspaces $\mathcal{M}(\vec{d})$ have a finite higher energy than the lowest energy state within $\mathcal{M}(\vec{0})$.

\begin{figure}[htb]
\centering
\includegraphics[height=0.38\linewidth]{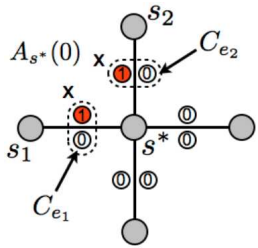}
\caption{A non-ground-state subspace $\mathcal{M}(\vec{d})$. $A_{s^{*}}(0)$ anti-commutes with two edge terms $C_{e_{1}}$ and $C_{e_{2}}$.
} 
\label{gap}
\end{figure}

We first consider a subspace $\mathcal{M}(\vec{d})$ defined by $\ket{\psi(\vec{d})} = A_{s^{*}}(0) \ket{\psi(\vec{0})}$ where $\vec{d}$ has non-zero components for two qubits acted on by $A_{s^{*}}(0)$, as shown in Fig.~\ref{gap}. We notice that $A_{s^{*}}(0)$ commutes with all terms except two edge terms $C_{e_{1}}$ and $C_{e_{2}}$. Therefore, solving $H_{gadget}$ inside $\mathcal{M}(\vec{d})$ is equivalent to solving 
\myequationn{A_{s^{*}}(0)H_{gadget}A_{s^{*}}(0)^{\dag} = H_{gadget} + V}
inside $\mathcal{M}(\vec{0})$, where $V = 2J(C_{e_1} + C_{e_2})$.

Below, we show that the lowest energy states for $H_{gadget} + V$ inside $\mathcal{M}(\vec{0})$ have finite higher energy than those of $H_{gadget}$ for appropriate choices of parameters $U$, $t$ and $J$. For simplicity of discussion, we neglect a constant correction resulting from plaquette term $H_{p}$ by writing $H_{gadget}=H_{hop}=\sum_{s}h_{s}$ inside $\mathcal{M}(\vec{0})$. Then, one may write $H_{gadget}'  = \sum_{s \not= \{ s^{*},s_{1},s_{2}\}} h_{s} + H^{*}$ with
\begin{align}
H^{*} = \sum_{s = \{ s^{*},s_{1},s_{2}\}} h_{s} + 2J(C_{e_{1}} + C_{e_{2}})\notag
\end{align}
where $s_{1}$ and $s^{*}$ are connected by $e_{1}$, and $e_{2}$ connects $s_{2}$, $s^{*}$ (Fig.~\ref{gap}).

Returning to the $\lambda$-representation, we note that all particles except $s^{*}, s_1, s_2$ are non-interacting and are governed under the same Hamiltonian $h_{s}$ as before. Let us denote the lowest energy eigenvalue of $h_{s}$ as $E_{0}$. Noting that $E_{0}$ is upper bounded by $-U$, it suffices to show that $H^{*}> - 3U>3E_{0}$ for the existence of an energy gap.

Let $H^{*}=H_{1}+H_{2}$ where $H_{1} = -t \sum_{s = \{ s^{*},s_{1},s_{2}\} } (D_{s}^{\dagger} + D_{s})$ and $H_{2}=-U\sum_{s = \{ s^{*},s_{1},s_{2}\} } |m_{s}=0\rangle \langle m_{s}=0| + 2J(C_{e_{1}}+C_{e_{2}})$. Since one cannot minimize $H_{1}$ and $H_{2}$ simultaneously, we obtain a lower bound for $H^{*}$ by finding minimal energy eigenvalues for $H_{1}$ and $H_{2}$ individually. One can verify that $H_{1} \geq -6t$ by directly finding energy eigenvalues of $H_{1}$. Similarly, one can verify that $H_{2} \geq \min(-3U+4J,-2U-4J)$. Here, let us choose $U$ and $J$ such that $J = U/8$, and $H_{2} \geq -\frac{5}{2} U$. $H_{gadget}'$ has a provably higher ground state energy than $H_{gadget}$ when $H^{*} > -6t - \frac{5U}{2} > -3U>3E_{0}$, so we simply set $U>12t$. This proof may be easily generalized to arbitrary $\mathcal{M}(\vec{d})$ when $U>16t$.

A drawback of this proof is that a small value of $t = U/16$ gives a weak constant gap for $h_s$ and thus the gap inside $\mathcal{M}(\vec{0})$ is  $\sim10^{-4} U$. Tighter analysis presented in appendix \ref{sec:bound} shows that when $J = 0.09 U $, $t = 0.375 U$, the system has a quite reasonable energy gap of $>0.075 U$ both inside and outside $\mathcal{M}(\vec{0})$.

\emph{Particle dimension---} In this construction, a gadget particle is four-dimensional, and a composite particle is eight-dimensional after removing the internal degree of freedom for $B_p$. This can be improved through defining a similar construction on a triangular lattice, leading to six-dimensional gadget particles and four-dimensional composite particles. In addition, a more elaborate construction reduces the dimension of the gadget particles to three while keeping the dimension of composite particles at four (see appendix \ref{sec:lessdim}).

\emph{Discussion---}
Our gadget construction can be generalized to the quantum double model, which may be universal for topological quantum computation, in a rather straightforward way (see appendix \ref{sec:double}). We expect that similar generalizations are possible for other interesting, but highly non-local topologically ordered Hamiltonians. In addition, our non-perturbative gadget may find use in adiabatic quantum computation and Hamiltonian complexity problems. 

In our construction, we have heavily taken advantage of the fact that the terms being simulated commute. Whether non-commuting terms can be simulated in this way remains an open question. Perhaps insights from related problems in theoretical computer science will prove fruitful, opening new connections. 

\emph{Conclusion---} In this paper, we argue the necessity of simulating topological quantum codes \emph{non-perturbatively}, and propose a model which is two-body, exactly solvable, and supports completely localized quasi-particle excitations. 
While our construction involves large particle dimensions and \quotes{gadgety} interaction terms which put it beyond modern experimental capabilities, there has recently been remarkable theoretical and experimental progress in engineering custom interactions between particles~\cite{Micheli06, Verstraete09, Weimer10}.
We hope our construction will provide a stepping stone towards physical realizability of topological quantum codes.

\emph{Acknowledgements---}
SAO is supported by NSF Grant No. DGE-0801525, {\em IGERT: Interdisciplinary Quantum Information Science and Engineering}. BY thanks Eddie Farhi and Peter Shor for support. BY is supported in part by DOE Grant No. DE-FG02-05ER41360 and by Nakajima Foundation. 


%

 \onecolumngrid
   \newpage
  \appendix

\section{Designing the shielding term}\label{sec:shield}

Here, we present an explicit procedure to obtain the shielding term $H_{shield}$ which cancels the edge term $H_{e}$ inside the ground state subspace $\mathcal{M}(\vec{0})$.
\begin{figure}[htb]
\centering
\includegraphics[width=0.5\linewidth]{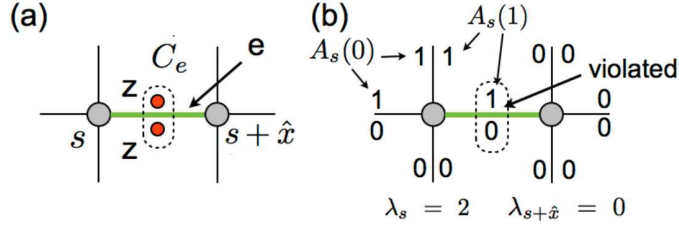}
\caption{Construction of the shielding term. (a) The horizontal edge. (b) A violation of an edge term $C_{e}$. 
} 
\label{shield}
\end{figure}

\textbf{Horizontal edge terms:}
Let us first consider an edge term $C_{e}$ for a horizontal edge $e$, connecting two stars $s$ and $s+ \hat{x}$ (see Fig.~\ref{shield}(a)). We represent $C_{e}$ in the \quotes{$\lambda$-representation} (see Eq.~(2) in the main article) inside the ground state subspace $\mathcal{M}(\vec{0})$. Since the edge term $C_{e}$ acts only on $|\lambda_{s}\rangle$ and $|\lambda_{s + \hat{x}}\rangle$ nontrivially, we will compute the diagonal elements $\langle \lambda_{hor}| C_{e}|\lambda_{hor}\rangle$ for $|\lambda_{hor} \rangle \equiv |\lambda_{s}\rangle \otimes |\lambda_{s+\hat{x}}\rangle$ where $\lambda_{s}, \lambda_{s+\hat{x}}=0,\ldots,7$. 

As an example, let us compute the diagonal element for $\lambda_{s}=2$ and $\lambda_{s+\hat{x}}=0$ (see Fig.~\ref{shield}(b)). Then, the edge term $C_{e}$ is violated since two qubits on the edge $e$ have different values, and $\langle \lambda_{hor}| C_{e}|\lambda_{hor}\rangle = -1$. On the other hand, for $\lambda_{s}=0$ and $\lambda_{s+\hat{x}}=0$, we have $\langle \lambda_{hor}| C_{e}|\lambda_{hor}\rangle = 1$ since both qubits on the edge $e$ are in $|0\rangle$. 

By repeating similar analyses for every pair of $\lambda_{s}$ and $\lambda_{s+\hat{x}}$, we have:
\begin{align}\notag
	  \langle \lambda_{hor}| C_{e}|\lambda_{hor}\rangle = \left\{
     \begin{array}{lr}
       -1 & : \lambda_{s}=2,6 \quad \mbox{and}\quad  \lambda_{s+\hat{x}}=0,4 \\
       \phantom{-}1 & :\lambda_{s}\not=2,6 \quad \mbox{and}\quad  \lambda_{s+\hat{x}}=0,4\\
       \phantom{-}1 & :\lambda_{s}=2,6 \quad \mbox{and}\quad  \lambda_{s+\hat{x}}\not=0,4\\
       -1 & :\lambda_{s}\not=2,6 \quad \mbox{and}\quad  \lambda_{s+\hat{x}}\not=0,4
     \end{array}
     \right.
\end{align}
which we can simplify as:
\begin{align}\notag
\langle \lambda_{hor}| C_{e}|\lambda_{hor}\rangle = 
 \paren{1 - 2 \delta_{\lambda_{s, 2}} - 2 \delta_{\lambda_{s}, 6}} \cdot   \paren{2 \delta_{\lambda_{s+\hat{x}}, 0} + 2 \delta_{\lambda_{s+\hat{x}}, 4}  -1}. \notag
      \end{align}
Noting that $m_s = \lambda_s \ (\mbox{mod}\ 4)$ $\forall s$ inside $\mathcal{M}(\vec{0})$, we see that $
\bra{\lambda_{hor}} C_{e}\ket{\lambda_{hor}}   = \bra{\lambda_{hor}}T_{\ell}(m_{s})T_{r}(m_{s+\hat{x}})\ket{\lambda_{hor}} $
within $\mathcal{M}(\vec{0})$ where 
\myequationn{
T_{\ell}(m) = 1-2 \delta_{m,2}, \quad T_{r}(m) = 2 \delta_{m,0} -1.}

Therefore, a shielding term $J \cdot T_{\ell}(m_{s})T_{r}(m_{s+\hat{x}})$ exactly cancels the two-body contribution arising from $C_{e}$ since $\langle \lambda_{hor}|- J \cdot C_{e}+ J \cdot T_{\ell}(m_{s})T_{r}(m_{s+\hat{x}})|\lambda_{hor}\rangle=0$ inside $\mathcal{M}(\vec{0})$.\\

\textbf{Vertical edge terms:}
Next, let us consider an edge term $C_{e}$ for a vertical edge $e$, connecting two stars $s$ and $s+ \hat{y}$. Then, for $\ket{\lambda_{ver}}=\ket{\lambda_{s}} \otimes\ket{\lambda_{s+\hat{y}}} $, one can verify that: 
\begin{align}\notag
\langle \lambda_{ver}| C_{e}|\lambda_{ver}\rangle  = 
 \paren{1 - 2\delta_{\lambda_{s},1}- 2\delta_{\lambda_{s}, 5}} \cdot \paren{1 - 2 \delta_{\lambda_{s+\hat{y}}, 3} - 2 \delta_{\lambda_{s+\hat{y}}, 7}}.\notag
\end{align}
Then, for a shielding term $ T_{d}(m_{s})T_{u}(m_{s+\hat{y}})$, we have $\langle \lambda_{ver}|- J \cdot C_{e} +J \cdot T_{d}(m_{s})T_{u}(m_{s+\hat{y}})|\lambda_{ver}\rangle=0$ inside $\mathcal{M}(\vec{0})$.\\

Therefore, $H_{e} + H_{shield} = 0$ inside $\mathcal{M}(\vec{0})$.

\newpage
\section{Improved bound on energy gap}\label{sec:bound}

In this section we improve the energy gap by improving the energy bound on non-ground-state subspaces $\mathcal{M}(\vec{d})$. Within any subspace, $B_{p} = \pm 1$, and when $B_{p} = -1$ the energy is raised without affecting other terms. Therefore, we need only consider non-ground-state subspaces where $B_{p} = 1$ and neglect a constant correction from $H_{p}$. As seen previously, solving $H_{gadget}$ inside $\mathcal{M}(\vec{d})$ is equivalent to solving $H_{gadget} + V$ inside $\mathcal{M}(\vec{0})$, where:
\myequationn{
V = 2J \paren{\sum_{e_{hor} \in \viol}C_{e_{hor}}     + \sum_{e_{ver} \in \viol}C_{e_{ver}}    }.
}
Here, $\viol$ contains all edges $e: C_{e} \ket{\psi(\vec{d})} = -1$. We note that within $\mathcal{M}(\vec{0})$, for all vertical and horizontal edges:
\begin{align*}
 C_{e_{hor}}  &= T_{\ell}(m_{s})T_{r}(m_{s+\hat{x}}) \geq  T_{\ell}(m_{s}) + T_{r}(m_{s+\hat{x}}) -1 \\ 
  C_{e_{ver}}  &= T_{d}(m_{s})T_{u}(m_{s+\hat{y}}) \geq  T_{d}(m_{s}) + T_{u}(m_{s+\hat{y}}) -1 .
\end{align*}
Therefore, a lower bound on $H_{gadget} + V'$, where 

\myequationn{
V' = &2J \paren{\sum_{e_{hor} \in \viol} \paren{T_{\ell}(m_{s}) + T_{r}(m_{s+\hat{x}}) -1 }   
  +\sum_{e_{ver} \in \viol}\paren{T_{d}(m_{s}) + T_{u}(m_{s+\hat{y}}) -1   }}
}
also serves as a lower bound on $H_{gadget} + V$. This is useful as $H_{gadget} + V'$ is one-body in the $\lambda$-representation. 

Organizing terms by stars instead of by edges, we find that: 
\myequationn{
H_{gadget} + V' = \sum_{s \notin \viol} h_{s} + \sum_{s^{*} \in \viol} h'_{s^{*}},
} where 
\myequationn{
h'_{s*}  =h_{s^{*}} + 
&2J \Big( a^{\ell}_{s^{*}}(T_{\ell}(m_{s^{*}}) - 1/2  ) +  a^{r}_{s^{*}}(T_{r}(m_{s^{*}}) - 1/2 ) +\\
&2J \Big( a^{d}_{s^{*}}(T_{d}(m_{s^{*}}) - 1/2 )+ a^{u}_{s^{*}}(T_{u}(m_{s^{*}}) - 1/2 )\Big).
}
The coefficients $a^{\ell}_{s^{*}}, a^{r}_{s^{*}}, a^{d}_{s^{*}}, a^{u}_{s^{*}} = 0, 1$ denote whether $C_{e} \ket{\psi(\vec{d})} = \pm \ket{\psi(\vec{d})}$ for the left, right, down, and up edges of $s^{*}$. The basic idea of our analysis is to find a lower bound for the energy of $h'_{s^*}$ for all $2^4-1 = 15$ cases where $a^{\ell}_{s^*}+a^{r}_{s^{*}} + a^{d}_{s^{*}} + a^{u}_{s^{*}}>0$.

 We can further tighten analysis by rewriting the above equation:
\myequationn{
h'_{s*}  =h_{s^{*}} + 
&2J \Big( a^{\ell}_{s^{*}}(T_{\ell}(m_{s^{*}}) - 1/2 - \beta_{\ell r} ) +
  a^{r}_{s^{*}}(T_{r}(m_{s^{*}}) - 1/2 +\beta_{\ell r}) \Big)  + \\
&2J \Big( a^{d}_{s^{*}}(T_{d}(m_{s^{*}}) - 1/2 -\beta_{du}) +
 a^{u}_{s^{*}}(T_{u}(m_{s^{*}}) - 1/2 +\beta_{du}) \Big).
}
$\beta_{\ell r}$ and $\beta_{d u}$ are constants to be optimized. This modification simply redistributes constant energy between stars, leaving the \emph{total} Hamiltonian unchanged; $\sum_{s^{*}} a^{\ell}_{s^{*}} = \sum_{s^{*}} a^{r}_{s^{*}}$ and $\sum_{s^{*}} a^{d}_{s^{*}} = \sum_{s^{*}} a^{u}_{s^{*}}$. We vary parameters to find an optimum at $J = 0.09 U $, $t = 0.375 U$, $\beta_{\ell r} = 0.25$, $\beta_{d u}  =0$. For these values, $h'_{s*} > E_{0} + 0.25 U \forall a^{\ell}_{s^{*}}+ a^{r}_{s^{*}}+ a^{d}_{s^{*}}+ a^{u}_{s^{*}} >0$. Recall that $E_{0}$ is the ground state energy of $h_s$. 

Since any non-ground-state subspace $\mathcal{M}(\vec{d})$ must have at least three stars $s^{*}$ for which this holds, the lowest energy of any state in $\mathcal{M}(\vec{d})$ is at least $0.075 U$ above the ground state energy. Likewise, at this value of $t$ the gap to a single \quotes{vortex} excitation in $h_{s}$ is $>0.0375U$. Since an even number of stars $s$ must be excited in this way, any excited state \emph{inside} $\mathcal{M}(\vec{0})$ must have energy at least $0.075U$ higher than the ground state energy. We combine these two bounds to prove a quite reasonable energy gap of at least $0.075 U$.


\section{Gadget for quantum double}\label{sec:double}

Here, we present a generalization of our gadget construction to the quantum double model. Our construction and discussion closely parallel that of the toric code, but are somewhat more complicated.  We begin by defining a modified version of the quantum double model that we will simulate through a two-body Hamiltonian. Consider an arbitrary finite group $G$ (which may be non-abelian), and consider a qudit with an orthogonal basis $\{ |z\rangle : z \in G \}$ whose dimensionality is $|G|$. We define the following group operations:
\myequationn{
      \begin{matrix} 
      L_{+}^{g}|z\> = |gz\> 			& \quad  L_{-}^{g}|z\> = |zg^{-1}\> \\
      T_{+}^{h}|z\> = \delta_{h,z} |z\> 	& \quad T_{-}^{h}|z\> = \delta_{h^{-1},z} |z\> \\
   \end{matrix}
}

\begin{figure}[h!]
\centering
\includegraphics[width=0.5\linewidth]{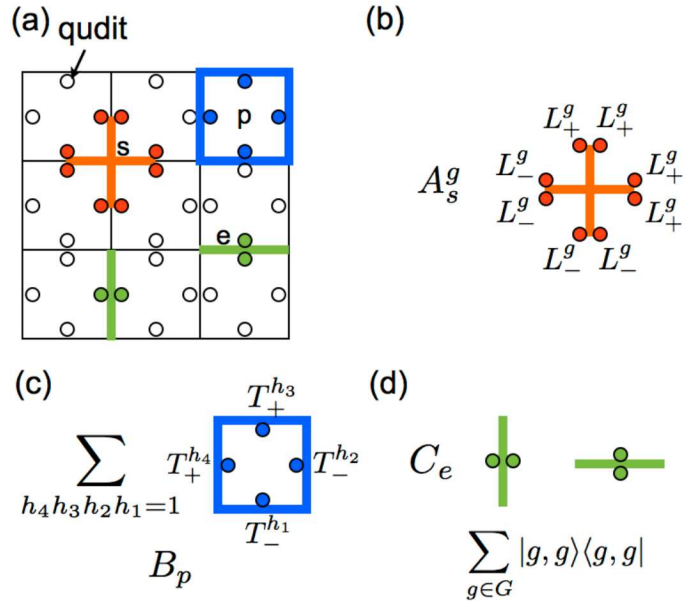}
\caption{(a) The modified quantum double model. (b) A star term. (c) A plaquette term. (d) An edge term.
} 
\label{double}
\end{figure}

Note that, while the group may be non-abelian, $L_{+}^{g}$ and $L_{-}^{h}$ nevertheless commute, as $(gz)h^{-1} = g(z h^{-1})$.
We consider a system of qudits defined on edges of a square lattice with periodic boundary conditions, where two qudits live on each edge in our construction (see Fig.~\ref{double}(a)). The system is governed by the Hamiltonian:
\begin{equation*}
\begin{split}
H =  H_{s} + H_{p} + H_{e}.
\end{split}
\end{equation*}

The star term is:
\begin{align*}
H_{s} = - J \sum_{s} \sum_{g\in \gen} A^{g}_{s} , 
\end{align*}
where $A^{g}_{s}$ is represented in Fig.~\ref{double}(b) and $\gen$ is a generating set of group $G$; any element in $G$ is some product of elements in $\gen$. The plaquette term is 
\begin{align*}
H_{p} = - J \sum_{p} B_{p}, 
\end{align*}
where $B_p$ projects onto the subspace where the clockwise product of qudit group elements in a plaquette is the identity, as represented in Fig.~\ref{double}(c). The edge term is 
\begin{align*}
H_{e} = - J \sum_{e} C_e,
\end{align*}
where $C_{e}$ projects onto the subspace where the two qudits on $e$ have the same group element, as represented in Fig.~\ref{double}(d). Again, all terms $A_{s}$, $B_{p}$, $C_{e}$ commute and can be minimized simultaneously. It can similarly be verified that the ground states of the original and modified quantum double model can be connected through generalized controlled-NOT gates between qudits on edges. Grouping four qudits in each plaquette into a single composite particle, $A^{g}_{s}$ becomes four-body, $B_p$ becomes one-body, and $C_e$ is two-body. Below, we will show how $A^{g}_{s}$ can similarly be simulated through two-body interactions.

\textbf{Gadget Hamiltonian:}
We again add a gadget particle to each star. The gadget particle again has a spin degree of freedom $m_{s} = 0,1,2,3$, but now additionally has a group element degree of freedom: $g_{s} \in \gen$ where $\gen$ is a generating set of the group $G$. Therefore, the gadget particle at position $s$ is described by $\ket{m_{s}, g_{s}}$, and has dimension of $4 |\gen|$. We replace the four-body star terms $A^{g}_{s}$ with two-body interaction terms:
\begin{equation*}
\begin{split}
&H_{gadget} = H_{p} + H_{e} + H_{hop} + H_{shield}  \\ 
&H_{p} = - J\sum_{p} B_p,\quad H_{e} = - J\sum_{e} C_e. 
\end{split}
\end{equation*}

\begin{figure}[htb!]
\centering
\includegraphics[width=.6 \linewidth]{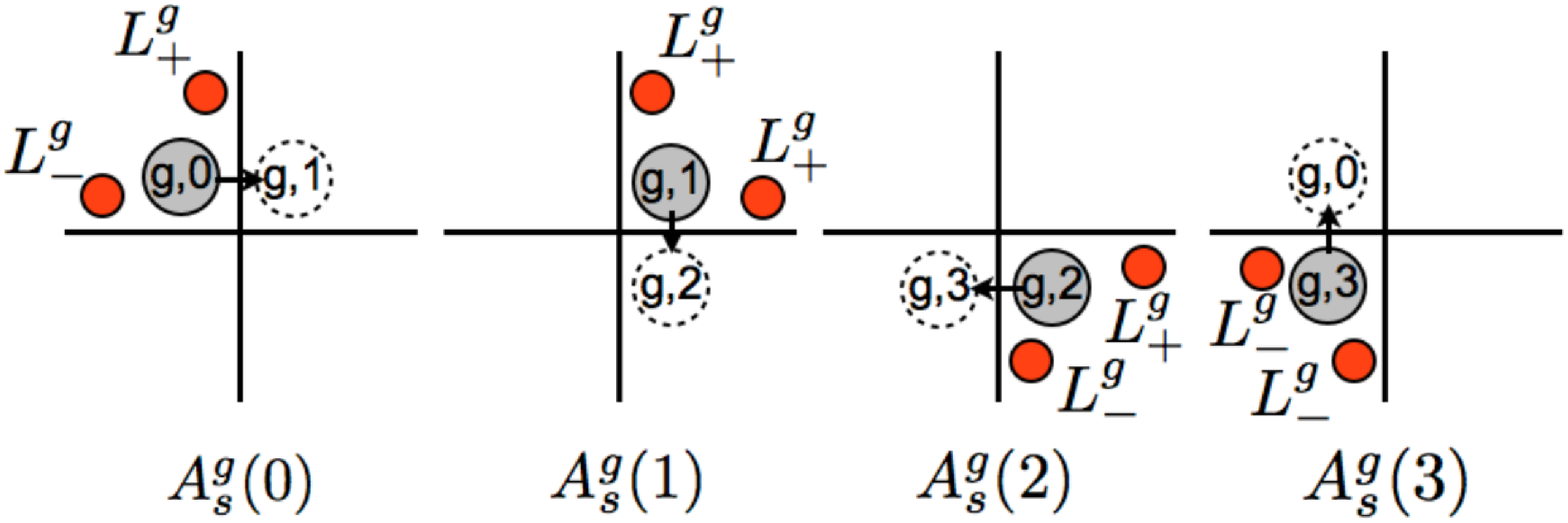}
\caption{The hopping term $\paren{D^{g}_{s}}^{\dag}$ for the quantum double model.
} 
\label{double_hopping}
\end{figure}

The hopping term is $H_{hop}=\sum_{s}h_{s}$ where
\begin{align}
\begin{split}
h_{s} &= -  U| m_{s}=0 \rangle\langle m_{s}=0|  -t \cdot Q_{s}-t \sum_{g \in \gen}\paren{D^{g}_{s} }^{\dagger} + D^{g}_{s}   \\ 
\paren{D^{g}_{s}}^{\dag} &=  \sum_{m_{s}=0,1,2,3} | m_{s}+1, g \rangle \langle m_{s}, g | \otimes A^{g}_{s}(m_{s}) \quad \mbox{(mod $4$)}\notag\\
Q_{s} &= \ \   \sum_{g, g' \in \gen}\ket{m_{s} =0 , g_{s} = g}\bra{m_{s} = 0, g_{s} = g'}.
\end{split}
\end{align}

Terms $A^{g}_{s}(m)$ are products of two $L_{\pm}^{g}$ operators as depicted in Fig. \ref{double_hopping}. Since $A_{s}^{g}(m)$ are one-body operators when plaquettes are viewed as composite particles, hopping terms are two-body. The hopping terms effectively induce star terms since $A_{s}^{g} = \prod_{m = 0}^{3} A_{s}^{g}(m)$.

The shielding term is: 
\myequationn{
H_{shield} = \frac{J}{4} &\sum_{s}  \bigg( \paren{1+T_{\ell}(m_{s})} \paren{1+T_{r}(m_{s+\hat{x}})}+
 \delta_{g_{s}g_{s+\hat{x}}} \paren{1-T_{\ell}(m_{s})} \paren{1-T_{r}(m_{s+\hat{x}})} \bigg) + \\
 \frac{J}{4} & \sum_{s}  \bigg( \paren{1+T_{d}(m_{s})} \paren{1+T_{u}(m_{s+\hat{y}})}+
 \delta_{g_{s}g_{s+\hat{y}}} \paren{1-T_{d}(m_{s})} \paren{1-T_{u}(m_{s+\hat{y}})} \bigg),
 }
which will decouple effective interactions to make the model exactly solvable. This shielding term may be derived in a similar way as was done for the toric code.

\textbf{Decomposition into subspaces:}
We can analogously decompose this gadget Hamiltonian into subspaces to help us solve it. We denote computational basis states whose gadget spin values are all $\ket{0}$s as 
\myequationn{\ket{\psi(\vec{g}, \vec{d})} =  \bigotimes_{s} \ket{ m_{s}=0, g_{s}}_{gadget} \otimes \ket{\vec{d}}_{qudit}.} 
We define the subspace $\mathcal{M}(\vec{d})$ such that it is spanned by all the states which can be reached from $\ket{\psi(\vec{g}, \vec{d})}$ by applying $\paren{D_{s}^{g_s}}^{\dag}$ for some $\vec{g}$:
\myequationn{
\mathcal{M}(\vec{d}) = \Bigg \< \ \prod_{s}\paren{\paren{D_{s}^{g_{s}}}^{\dag}}^{\lambda_{s}}   \ket{\psi(\vec{g}, \vec{d})}, \   \forall \vec{g}, \vec{\lambda}   \  \Bigg\>,}
and can verify that $\mathcal{M}(\vec{d})$ is an invariant subspace of $H_{gadget}$. Then, we can solve the gadget Hamiltonian inside each subspace $\mathcal{M}(\vec{d})$ independently. 

\textbf{Ground state subspace:}
We solve for the lowest energy state inside $\mathcal{M}(\vec{0})$ where the identity element in the group $G$ is denoted by $0$. We note that $B_{p} = 1$ inside $\mathcal{M}(\vec{0})$ and thus need not be considered. Denoting the total number of stars as $N$, we may view $\mathcal{M}(\vec{0})$ as the Hilbert space of $N$ particles, using a somewhat more complicated \quotes{$\lambda$-representation}:
\myequationn{
\bigotimes_{s} \ket{\lambda_{s}, g_{s}, f_{s}} = \prod_{s}\paren{\paren{D^{g_{s}}_{s}}^{\dag}}^{\lambda_{s}} A^{f_s}_{s}  \ket{\psi(\vec{g}, \vec{0}) }.
}
We note that this representation is redundant by seeing:
 \myequationn{ \paren{(D_{s}^{g})^{{\dag}} }^{4} \ket{\psi(\vec{g}, \vec{d)}} = A^{g}_{s} \ket{\psi(\vec{g}, \vec{d)}}}
Translating into the $\lambda$-representation, the above redundancy becomes:
\myequation{\ket{\lambda_{s}+4, g_{s}, f_{s}} =  \ket{\lambda_{s}, g_{s}, \ g_{s} \cdot f_{s}}. \label{doubletruncate}}
This allows us limit ourselves to $\lambda_{s} = 0, 1,2,3$ in the $\lambda$-representation, giving each \quotes{particle} a Hilbert space of finite dimension $4 |\gen| |G|$.

We can confirm that, within $\mathcal{M}(\vec{0})$,  $H_{shield}+ H_{e} = 0$. Therefore, in the \quotes{$\lambda$-representation}, $H_{gadget}$ acts as a one-body Hamiltonian $H_{gadget}  = const + \sum h_{s}$, where:
\myequationn{
h_{s} = -U   \sum_{g_s, \ f_s}&\underbrace{\proj{\lambda_{s} = 0, g_s, f_s}}_{\delta_{m_s}}\\
-t \sum_{\lambda_{s} = 0,1,2,3} \ \sum_{ g_s, \ f_{s}} &\underbrace{\ket{\lambda_{s}+1, g_{s}, f_{s}}\bra{\lambda_{s}, g_{s}, f_{s}} + h.c.}_{(D^{g_s}_{s})^{\dag} + D^{g_s}_{s}} \\
-t \sum_{g_s, \ g'_s, \  f_s} &\underbrace{ \ket{\lambda_{s} = 0, g_{s}, f_{s}}\bra{\lambda_{s} = 0, g'_{s}, f_{s}}}_{Q_s},
}
and Eq.~\eqref{doubletruncate} is implicit. 

We can write the lowest energy state inside $\mathcal{M}(\vec{0})$ as $\ket{\psi_{GS}(\vec{0}) }= \bigotimes_{s} \ket{\alpha_{0}}$, where $\ket{\alpha_{0}} = \sum_{\lambda, g, f} \alpha_{0}(\lambda) \ket{\lambda, g, f}$,
and returning from the $\lambda$-representation, we write it as:
\begin{equation}
\begin{split}
|\psi_{GS}(\vec{0})\rangle &=  \sum_{\vec{g}} \prod_{s}\sum_{f} A^{f}_{s} \sum_{\lambda=0}^{3}  \alpha_{0}(\lambda) \paren{\paren{D_{s}^{g_s}}^{\dag}}^{\lambda} \ket{\psi(\vec{g}, \vec{0})}. \notag
\end{split}
\end{equation}
One can verify that $|\psi_{GS}(\vec{0})\rangle$ is the ground state of the gadget Hamiltonian, and the energy gap may be proven in the same manner as was done for the toric code.

\textbf{Unitary Connection: }The ground state $|\psi_{GS}(\vec{0})\rangle$ is connected to the ground state of the modified quantum double model through the following local unitary transformation: $U = \prod_{s} U_s$ where:
\begin{align*}
U_{s} \equiv \sum_{g_s \in \gen} \sum_{m_{s}=0}^{3} \proj{m_s, g_s} \prod_{m < m_{s}}A^{g}_{s}(m)^{\dag}. \label{eq:unitary}
\end{align*}
In particular, we have $U|\psi_{GS}(\vec{0})\rangle = |\tilde{\alpha_{0}} \rangle_{gadget}^{\otimes N} \otimes |\psi_{QD}(\vec{0})\rangle_{qudit}$ where $\ket{\tilde{\alpha_{0}}}= \sum_{g \in \gen} \sum_{m=0}^{3}\alpha_{0}(m) \ket{m, g}$, and 
$|\psi_{QD}(\vec{0})\rangle_{qudit} = \prod_{s} \paren{\sum_{f \in G} A^{f}_{s} }\ket{\vec{0}} $ which is a ground state of the modified quantum double. In fact, $U$ maps each ground state of the gadget Hamiltonian to a corresponding ground state of the modified quantum double.

\textbf{Particle dimension:} The particle dimension can be somewhat reduced in the same way as in the toric code; defining a similar construction on a triangular lattice and removing the degree of freedom $B_{p}$ gives us plaquettes with a $|G|^{2}$-dimensional Hilbert space and gadget particles with a $6|\gen|$-dimensional Hilbert space. A modification similar to the one presented in appendix \ref{sec:lessdim} can reduce the dimension of the gadget particles to $1 + 2|\gen|$ while keeping the dimension of the composite particles the same.


\section{Reducing particle dimension}\label{sec:lessdim}

\begin{figure}[htb]
\centering
\includegraphics[width=0.5\linewidth]{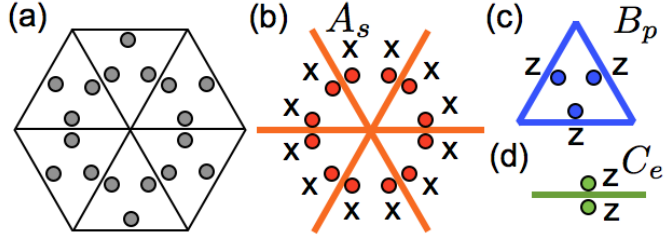}
\caption{(a) Construction of the modified toric code on a triangular lattice. (b) A star term. (c) A plaquette term. (d) An edge term.} 
\label{fig_reducing}
\end{figure}

Here we present an extension of our gadget construction which has three-dimensional gadget particles and four-dimensional composite particles. Our construction and discussion closely parallel those of the main paper, but are somewhat more complicated. We define a modified version of the toric code on a \emph{triangular} lattice with periodic boundary conditions as represented in Fig.~\ref{fig_reducing}, governed by the Hamiltonian:

\begin{equation*}
\begin{split}
&H = - J \sum_{s} A_{s} - J\sum_{p} B_{p} - J\sum_{e} C_{e} \\
&A_{s} = \prod_{j \in s} X_{j}, \quad B_{p} = \prod_{j \in p} Z_{j}, \quad C_{e} = \prod_{j \in e} Z_{j},
\end{split}
\end{equation*}

We group the three qubits in each plaquette into a single composite particle with an eight-dimensional space. While $B_{p}$ becomes one-body, and $C_{e}$ is two-body through this grouping, the star term $A_{s}$ is only reduced to six-body, and must be simulated using two-body terms involving gadget particles. 

\begin{figure}[htb]
\centering
\includegraphics[width=0.70\linewidth]{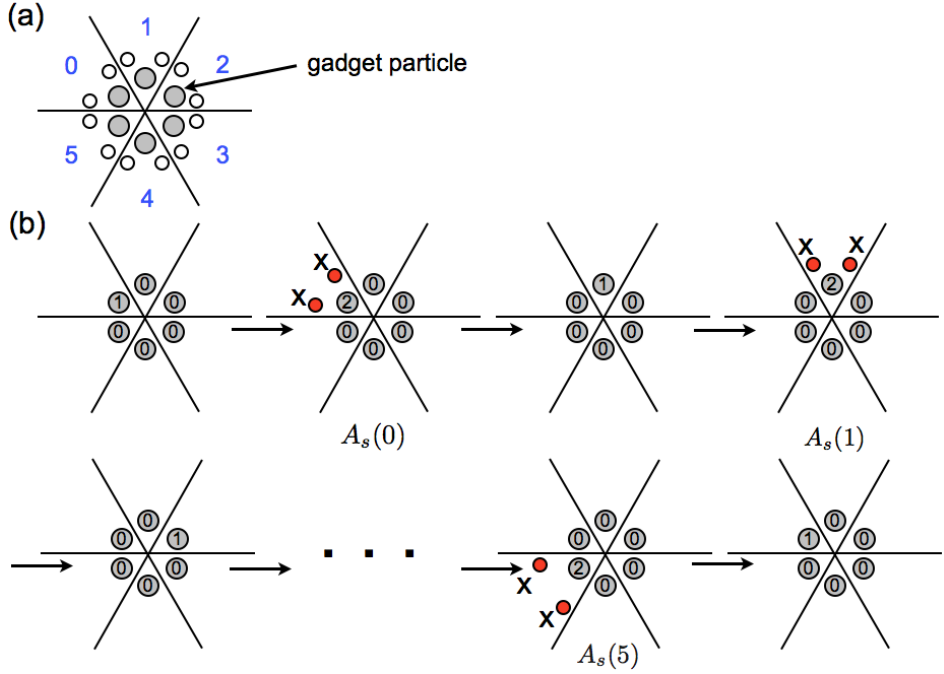}
\caption{Construction of the hopping term $D^{\dag}_{s}$. (a) Gadget particles live at corners of stars $s_{i}$. Labeling of corners $i$ is shown (blue online). Qubits live at edges.  (b) Terms $A_{s}(m)$ that are tensor products of two Pauli $X$ operators. Each term acts on two qubits (red online), depending on spin values of gadget particles. } 
\label{fig_reducing_gadget}
\end{figure}

\textbf{Gadget Hamiltonian:} This time we add \emph{six} gadget particles to each star $s$, one at each corner $i= 0 \ldots 5$. Each gadget particle $s_{i}$ has \emph{three} possible spin values, $m_{s_{i}} = 0, 1,2$. We replace the six-body star term $A_{s}$ with two-body terms $H_{hop}$ and $H_{shield}$ which involve the gadget particles:
\begin{equation}
\begin{split}
&H_{gadget} = H_{p} + H_{e} + H_{hop} + H_{shield}  \\ 
&H_{p} = - J\sum_{p} B_{p},\quad H_{e} = - J\sum_{e} C_{e}. \label{eq:gadget}
\end{split}
\end{equation}
The hopping term is $H_{hop}=\sum_{s}h_{s}$ where
\begin{equation}
\begin{split}
h_{s} &= -  U \proj{m_{s_{1}} = 1} - t  \left( D^{\dagger}_{s}  + D_{s} \right) +{R} \cdot \paren{1-n_{s}}^{2}\\
D_{s}^{\dagger} &= \sum_{i = 0}^{5} \ket{m_{s_{i}} = 2}\bra{m_{s_{i}} = 1} \otimes A_{s}(i) + \ket{m_{s_{i}} = 0} \bra{m_{s_{i}} = 2} \otimes  \ket{m_{s_{i+1}} = 1} \bra{m_{s_{i+1}} = 0}\\
n_{s} &= \sum_{i = 0}^{5} \paren{1 - \proj{m_{s_{i}} = 0}}
 \notag
\end{split}
\end{equation}
where $U$,  $t$, and {R} are some positive constants, and $m_{s_{i}}$ represents the spin value of the $i$th gadget particle at star $s$. Terms $A_{s}(i)$ are products of two Pauli $X$ operators as depicted in Fig.~\ref{fig_reducing_gadget}(b). Since $A_{s}(i)$ are one-body operators when qubits in a plaquette are viewed as a composite particle, hopping terms are two-body.
We note that $n_{s}$, the number of nonzero gadget particles at star $s$, is a good quantum number, and by choosing a sufficiently positive value of {R}, we restrict the ground space and low-energy excitations to the subspace where $n_{s} = 1$. In this subspace, $D_{s}^{\dagger}$ is unitary and the hopping term will effectively induce star terms $A_{s}$,  since $A_{s} = \paren{D_{s}^{\dagger}}^{12}$ as seen in Fig. \ref{fig_reducing_gadget}(b). 

The shielding term $H_{shield}$ consists of two-body interactions between gadget particles:
\begin{align}
H_{shield} =  J \sum_{s} \paren{T_{-\hat{u}_{0}}(s)T_{\hat{u}_{0}}(s+ \hat{u}_{0}) + T_{-\hat{u}_{1}}(s)T_{\hat{u}_{1}}(s+ \hat{u}_{1}) + T_{-\hat{u}_{2}}(s)T_{\hat{u}_{2}}(s+ \hat{u}_{2})   } \notag
\end{align}
\begin{align*}
T_{-\hat{u}_{0}}(s) = 1 - 2 \paren{\proj{m_{s_{2}} = 2} + \proj{m_{s_{3}} = 1}}, \quad   T_{\hat{u}_{0}}(s) = 2 \paren{\proj{m_{s_{0}} = 1} + \proj{m_{s_{5}} = 2}} -1, \\
T_{-\hat{u}_{1}}(s) = 1 - 2 \paren{\proj{m_{s_{1}} = 2} + \proj{m_{s_{2}} = 1}}, \quad   T_{\hat{u}_{1}}(s) = 1 -2 \paren{\proj{m_{s_{4}} = 2} + \proj{m_{s_{5}} = 1}}, \\
T_{-\hat{u}_{2}}(s) = 1 - 2 \paren{\proj{m_{s_{0}} = 2} + \proj{m_{s_{1}} = 1}}, \quad   T_{\hat{u}_{2}}(s) = 1 -2 \paren{\proj{m_{s_{3}} = 2} + \proj{m_{s_{4}} = 1}},
\end{align*}
where $s+\hat{u}_{j}$ are unit translations of a ``star'' $s$ to its nearest neighbors in the directions $\hat{u}_{0} = \hat{x}, \hat{u}_{1} = \frac{{\hat{x} + \sqrt{3}\hat{y}}}{2},  \hat{u}_{2} = \frac{{- \hat{x} + \sqrt{3}\hat{y}}}{2}$. As we will see below, this choice of the shielding term decouples effective interactions between neighboring gadget particles, making the model exactly solvable.

\textbf{Decomposition into subspaces:} Now, we solve the gadget Hamiltonian in Eq.~(\ref{eq:gadget}) inside the subspace $n_{s} = 1$. It is convenient to further decompose this space into  subspaces. Let us denote computational basis states whose gadget particles are at \quotes{rest} as: $|\psi(\vec{d})\rangle =  |\vec{\mathfrak{0}}\rangle_{gadget}  \otimes |\vec{d} \rangle_{qubit}$ where:  $\ket{\vec{\mathfrak{0}}}_{gadget} =\bigotimes_{s} \paren{ \ket{m_{s_{0}} = 1} \otimes_{i = 1}^{5} \ket{m_{s_{1}} = 0} }$. Again, $|\vec{d}\rangle_{qubit}$ represents spin values $\ket{d_{j}}$ for qubits, and we define the subspace $\mathcal{M}(\vec{d})$  to be spanned by all the states which can be reached from $|\psi(\vec{d})\rangle$ by applying $D_{s}^{\dagger}$:
\begin{align}
\mathcal{M}(\vec{d}) = \big \langle \prod_{s} (D_{s}^{\dagger})^{\lambda_{s}}|\psi(\vec{d})\rangle, \ \forall \lambda_{s} \big \rangle.\notag
\end{align}
We can verify that $\mathcal{M}(\vec{d})$ is an invariant subspace of $H_{gadget}$, and solve the gadget Hamiltonian inside each subspace $\mathcal{M}(\vec{d})$ independently.

\textbf{Ground state subspace:}
We solve for the ground state inside  $\mathcal{M}(\vec{0})$. We note that inside $\mathcal{M}(\vec{0})$, $B_{p} = 1$, and thus plaquette terms need not be considered. Again, we use the $\lambda$-representation:
\begin{equation}
|\vec{\lambda}\rangle = \bigotimes_{s} \ket{\lambda_{s}} =  \prod_{s} (D_{s}^{\dagger})^{\lambda_{s}}|\psi(\vec{0})\rangle. \label{eq:lambda}
\end{equation}
Noting that  $(D_{s}^{\dagger})^{12} = A_{s},  \quad (D_{s}^{\dagger})^{24}=  I$, these particles can be considered to have twenty-four-dimensional Hilbert spaces, $\lambda_{s} = 0, \ldots ,23$. In the $\lambda$-representation, the hopping term $H_{hop}$ can be written as a \emph{one-body} Hamiltonian: $H_{hop}=\sum_{s}h_{s}$ where:
\myequationn{
h_{s} =& - U \big( | \lambda_{s}=0  \rangle\langle \lambda_{s}=0| + | \lambda_{s}=12 \rangle\langle \lambda_{s}=12| \big)\\ 
&- t \sum_{\lambda_{s}=0}^{23} \big(|\lambda_{s} + 1 \rangle \langle \lambda_{s}|  + h.c \big) \qquad \text{ (mod $24$)}. \label{eq:hopping}
}
While edge terms $C_{e}$ are not one-body, inside $\mathcal{M}(\vec{0})$ they are exactly cancelled by the shielding term; $H_{shield} + H_{e} = 0$. 
This is because $C_{e} =  T_{-\hat{u}_{j}}(s)T_{\hat{u}_{j}}(m_{s+\hat{u}_{j}})$, where $j = 0, 1, 2$ for horizontal, upper-right diagonal, and upper-left diagonal edges respectively, which connect $s$ and $s+\hat{u}_{j}$. Because of this exact cancellation, the gadget Hamiltonian is one-body in the \quotes{$\lambda$-representation}: $H_{gadget}= \mbox{const}+\sum_{s}h_{s}$. 

Because of this, all energy eigenstates inside $\mathcal{M}(\vec{0})$ can be written in the tensor product form $\ket{\vec{\alpha}} = \bigotimes_{s} \ket{\alpha_{s}}$ where $\ket{\alpha_{s}} = \sum_{\lambda_{s}} \alpha_{s}(\lambda_s) \ket{\lambda_s}$. The lowest energy state is $\ket{\psi_{GS}(\vec{0})} = \bigotimes_{s}\ket{\alpha_0}$,  therefore, returning from the $\lambda$-representation, we can write the ground state as:
\begin{equation*}
\begin{split}
|\psi_{GS}(\vec{0})\rangle &= \prod_{s}\sum_{\lambda=0}^{23}\alpha_{0}(\lambda)(D_{s}^{\dagger})^{\lambda}|\psi(\vec{0})\rangle .
\end{split}
\end{equation*}

As with the construction in the main section, this model has a finite energy gap, four ground states connected to those of the original toric code by local unitary transformations, and supports completely localized quasi-particle excitations; all of these may be proved in analogous ways. \\

By removing the internal degree of freedom $B_{p}$ from the composite particles, we arrive at an exactly solvable model involving only four-dimensional (spin-$3/2$) composite particles and three-dimensional (spin-$1$) gadget particles.



\begin{thebibliography}{16}
\bibitem [{\citenamefont {OurFootnote}\ \emph {et~al.}(2010)  }]{OurFootnote}%
  \BibitemOpen
  \bibfield  {author} {Note that the $\lambda$-representation is doubly redundant, as $\ket{\vec{0}} = \prod A_{s} \ket{\vec{0}} = \ket{\vec{4}}$. This is consistent with the required anyonic statistics}\BibitemShut {NoStop}%
\end{thebibliography}
\end{document}